\documentstyle[aps,multicol,epsf]{revtex}
\begin{document}
\draft
\title{Effect of Doping-induced Disorder on Transition Temperature 
in High $T_c$ Cuprates}

\author{Hae-Young Kee}
\address{ Department of Physics, University of California,
Los Angeles, CA 90095}

\date{\today}
\maketitle

\begin{abstract}
We study the effect of disorder induced by doping 
on the transition temperature in high $T_c$ cuprates.
Since the impurity lies between the CuO$_2$ planes,
the momentum transfer is restricted by  small 
angle with a range  of order of $1/(d k_F)$, where $k_F$ is the Fermi
momentum and $d$ is the distance between the plane and the impurity.
We find that the leading correction on the transition temperature
in this case is in cubic order of $1/(d k_F)$,
while the single particle scattering rate is linearly proportional
to $1/(d k_F)$.
Therefore, we conclude that the disorder induced by doping has
rather small effect on $T_c$, while it could give large
single particle scattering rate.
We find that the reduction of $T_c$ is about $3-10 \%$,
using the single particle scattering
rate ($150-450 K$) observed in the recent angle-resolved-
photoemission-spectroscopy data on optimally doped 
Bi2212.
The consequences of the small angle scattering in other physical quantities
such as transport are also discussed.

\end{abstract}


\begin{multicols}{2}

A recent angle-resolved-photoemission-spectroscopy (ARPES) experiment
on Bi2212 shows
temperature independent single particle scattering rate
of about $150 -450 K$.\cite{valla2}
The scattering rate extracted to zero temperature 
shown in  the ARPES in the normal state
has been one of the puzzling issues in high $T_c$ cuprates.
Assuming that impurity
is responsible for such a large elastic scattering rate
in ARPES data,
one should be able to answer the following questions:
1. Why there is no substantial pair-breaking 
effect due to impurity.
2. Why  the transport scattering rate estimated from
optical conductivity\cite{quijada}  or residual resistivity
shows  anomalously small effect of impurity.
P. W. Anderson argued in a recent paper that this should be a 
clear evidence of the existence of ``quantum protectorate''
in high $T_c$ cuprates because, in conventional d-wave
superconductors, these impurities should act as strong pair-breakers
and lower $T_c$ down to zero temperature.\cite{pw}
Indeed, the impurity effect in the unitary limit on the d-wave 
superconductors was studied in \cite{maki1} and it was shown 
that the critical scattering rate which reduces $T_c$ down to 
zero temperature is $0.88 T_c$.

In order to get further insights in this issue,
let us recall some of the basic features of high $T_c$ cuprates.
One of the most important features of these materials
is that the superconducting state occurs only away from half filling.
It is clear that in any well-prepared cuprate sample 
there are always impurities coming from non-stoichiometry.
Naively, then,  one may conclude that
the doping-induced impurities\cite{note}
 should act as strong pair-breakers and reduce 
$T_c$ substantially since the superconducting state has 
d-wave symmetry.
However, in experiments, 
these impurities do not act as  strong pair-breakers.
Therefore, it is important to understand why they
have small effect on $T_c$ and also on the transport scattering rate,
but large effect on the single particle scattering rate.

In this paper, we investigate how the doping-induced impurities 
act on the single particle scattering rate and
the transition temperature $T_c$ based on the conventional
d-wave BCS picture.
We find that the reduction of $T_c$ due to small angle scattering
is rather small which shows the same effect as
the transport scattering rate, but not the single particle 
scattering rate.
Using the single particle scattering rate( $150 - 450 K$) observed in ARPES 
we find that 
the reduction of the transition temperature
is about  $3 \sim 10 \% $ for $a/d = 1/4 \sim 1/2$ where
$a$ is the lattice constant and $d$ is the distance between the impurity
and CuO$_2$ plane.
On the other hand, given the impurity scattering rate extracted
from the ARPES data\cite{valla2},
$T_c$ could be reduced down to zero temperature in the momentum-independent
weak scattering (Born) limit.

There has been a clear evidence that,
in optimally doped Bi2212, the imaginary part of the self
energy along the nodal direction 
has a linear dependence on temperature, independent
of energy for small binding energy, and a linear dependence
on energy, independent of temperature for large binding energy.
\cite{valla1}
The linear dependence of the imaginary part of self energy, Max(T, energy) 
is consistent with
the prediction made based on a phenomenological characterization
of the high $T_c$ materials as the marginal Fermi liquid.\cite{varma}
It has been also shown that within resolution limit, 
the temperature independent scattering rate is about $150 K $ at the nodal
direction and increases as one moves
away from the nodal direction.
The temperature independent offset was explained by 
taking into account
the small angle scattering due to the static impurities 
in the sample which are naturally induced by the doping of
Sr (LSCO) and O (YBCO, BSCCO).\cite{varma2}
Furthermore, the idea of the angular dependence
of the scattering rate around the Fermi surface
was proposed by considering the angular dependence
of the local density of states.\cite{varma2}

The above discussion suggests an investigation of
the effect of small angle scattering on the transition temperature.
Let us  consider the effect of doping in high $T_c$ cuprates.
The major effect of doping is, of course, changing
the concentration of the charge carriers.
Understanding the motion of the holes induced by doping in the 
antiferromagnetic background has been a long 
standing issue that is certainly related to the
mechanism of superconductivity in the cuprates, which 
we do not address in this paper.
On the other hand, the doping (say, by Sr in LSCO)
introduces impurities between the CuO$_2$ planes. 
Therefore, we can expect that the effect of this kind of 
disorder can be represented by
a smooth elastic scattering potential.
The following form of the  potential scattering can faithfully
represent the effect of disorder induced by doping.
\begin{equation}
u({\bf r}-{\bf r}^{\prime}) = \frac{u_0}{r^2+d^2},
\label{scattering}
\end{equation}
where $r= |{\bf r}-{\bf r}^{\prime}|$ and $u_0$ is the amplitude
of the scattering potential.

In order to represent the impurity scattering in the momentum space, 
we need to take a model Fermi surface.
We consider the simplest circular Fermi surface with the origin located
at $(0,0)$, where $a$ is
the lattice spacing.
The Fermi momentum measured from $(0,0)$
depends on the doping concentration.
For example, in the case of the optimally doped Bi2212, it is
reasonable to take $k_F \sim \frac{4\pi}{5a}$.\cite{valla2}
%

%
%
Taking a Fourier transform of the elastic potential, Eq. (\ref{scattering}),
with the model Fermi surface, we find that 
the scattering potential is represented by the modified Bessel
function of the second kind, 
$ K_0 \left( d |{\bf k}-{{\bf k}^{\prime}}| \right)$.
Therefore, 
it is reasonable to assume that  the scattering potential is
finite only for a range of small angles and can be described as
\begin{eqnarray}
u({\bf k}-{\bf k}^{\prime}) &= &  \cases{
                     u_0  &  ${\rm for} \; \phi-\phi^{\prime} < \theta_c$ \cr
                              0   & ${\rm otherwise}$ \cr },
\label{potential}
\end{eqnarray}
where
$\theta_c = \frac{1}{d k_F}$ and $\phi$ is the angle measured
from the $k_x$-axis.
It is important to notice that the small momentum transfer
(small angle scattering) is dominant in the scattering process.
The change of the angle due to the scattering is restricted by
$\frac{1}{d k_F}$.
Therefore, our analysis on the effect of the impurity on $T_c$
has to be distinguished from momentum-independent scattering, either
Born (weak) or unitary (strong) limit.
%
It can also be shown that the momentum transfer due to impurity
can be even smaller if the scattering potential is short-ranged
than the interaction form of Eq. (\ref{scattering}).


Let us consider the effect of the impurity potential, 
Eq. (\ref{potential}).
Since the effect of the impurity manifests itself through
the self energy, let us first compute the self energy
renormalized by the impurity scattering.
Assuming that the superconducting state is described by the d-wave
BCS superconductivity, we obtain the self energy using the
standard method described in\cite{maki2}.
The self energy is written as
\begin{equation}
\Sigma({\bf k}, {\tilde \omega})
= -\frac{n_i u_0^2}{2} \int^{\phi+\theta_c}_{\phi} d\phi^{\prime}
N(\phi^{\prime}) \frac{i {\tilde \omega} +
{\tilde \Delta}_0 \cos{(2 \phi^{\prime})} \rho_1 }
{\sqrt{{\tilde \omega}^2+{\tilde \Delta}_0^2 \cos^2{(2\phi^{\prime})}} },
\label{self}
\end{equation}
where $N(\phi)$ is the local density of states given by
$N(\phi) = \frac{m}{(2 \pi)}$.
Here $n_i$ is the impurity concentration and $\rho_i$ are the Pauli matrices
in the particle-hole space.
The dispersion of the quasiparticle 
in the normal state behaves as $\xi_{{\bf k}} = v_F p$,
where  $v_F$ is the Fermi velocity and
$p = |{\bf k} - {\bf k}_F|$ to obtain Eq. (\ref{self}).
The scattering rate  in the normal state
with our model Fermi surface
can be easily obtained by taking ${\tilde \Delta}_0=0$ 
and it is given by 
\begin{equation}
\frac{1}{\tau} (\phi)  = \frac{n_i u_0^2}{2} N(\phi) \theta_c.
\label{rate}
\end{equation}
As one can see, the scattering rate is proportional 
to the local density of states $N(\phi)$, multiflied by $\theta_c$.
Therefore, the scattering rate is proportional to the inverse of
the Fermi velocity, $1/v_F$.
Since our model Fermi surface is circular, the Fermi velocity
is independent of the angle $\phi$ and so is the single particle
scattering rate.

The angle dependence of
the scattering rate observed by ARPES\cite{valla2}
in the normal state(${\tilde \Delta}_0=0$) should
originate from the angle dependence of local density of
states through the Fermi velocity on the actual Fermi surface.
This can be seen through the simple analysis as follows.
Considering the actual Fermi surface shown in \cite{valla2},
we find that the Fermi momentum depends on the angle $\phi$
around the Fermi surface.
The doping-induced impurity potential, 
Eq. (\ref{scattering}) will lead to the same
result that the scattering is finite only for small range of
the angle, $\theta_c$,  which is now restricted by
$1/[d \sqrt{k_F^2+ (dk_F/d\phi)^2}]$.
On the other hand, the local density of states will be given by
$N(\phi) = m^* \sqrt{k_F^2+ (dk_F/d\phi)^2}/(2 \pi k_F)$,
where $m^*$ is the effective mass.
Using the fact that the scattering is finite only for small range of 
the angle around the Fermi surface,
we can expand the integrand of Eq. (\ref{self})
up to linear order in $\theta_c$ in the normal state(${\tilde \Delta}=0$).
Then the single particle scattering rate is given by
\begin{equation}
\frac{1}{\tau}(\phi) = \frac{n_i u_0^2}{2 d} \frac{1}{v_F(\phi)} 
\end{equation}
Therefore, the single particle scattering rate depends on
the angle around the Fermi surface through
the angular dependence of the Fermi velocity at each point
on the Fermi surface.
Although, the angular dependence of the single particle scattering rate
itself is an interesting issue and was used to explain the line shape
of the ARPES in \cite{varma}, 
we will show  below that
our analysis of the effect of small angle scattering on transition
temperature, qualitatively, does not depend on  whether the scattering
rate depends on angle or not.
We will keep the angle dependence of the local density
of states to make our analysis more general.
But we show below that the reduction of $T_c$ 
within our model Fermi surface
by taking $N(\phi)$ = constant is comparable with the result
obtained with the angle dependent scattering rate.

Let us proceed to determine the reduction
of the transition temperature due to impurity 
using the  method described in \cite{maki2}.
We first
consider the transition temperature in the absence of impurity,
which is obtained  by taking $\Delta_0 ={\tilde \Delta}_0 \rightarrow 0$
in the the gap equation,
\begin{equation}
\Delta_0= 8 V_0 T \sum_{n=1}^{\infty}  \int^{\pi/2}_{0} 
d\phi N(\phi) \frac{{\tilde \Delta}_0 \cos^2{(2\phi)} }
{\sqrt{{\tilde \omega}^2_n +{\tilde \Delta}^2_0 \cos^2{(2\phi)} } }.
\label{gap}
\end{equation}
One can see that the transition temperature $T_{c0}$ in the absence
of  impurity is given by
\begin{equation}
T_{c0}= \frac{2\gamma \omega_c}{\pi} \exp{(-1/[ 8 V_0 
 \int^{\pi/2}_{0} \frac{d\phi}{2\pi}
 N(\phi) \cos^2{(2\phi)} ] ) },
\end{equation}
where $\omega_c$ is the cut-off frequency determined
by the effective attractive interaction strength and $\gamma=1.768$.

In the presence of impurity, one needs to compute the ratio 
$\frac{ {\tilde \Delta}_0 }{{\tilde \omega}}$, which can be obtained
by using Eq. (\ref{self}) and the Dyson equation. 
One can show that the gap equation can be written as
\begin{eqnarray}
1 & = & 4 V_0 T_c \sum_{n=1} \int^{\pi/2}_{0} d\phi
\left[
\frac{N(\phi) \cos^2{(2\phi)}}{\omega_n} - n_i u_0^2 \right.
\nonumber\\ 
& \times & \left. \frac{N(\phi)^2 \cos^2{(2\phi)} 
+ N(\phi) \frac{d N(\phi)}{d\phi} \cos{(2\phi)} \sin{(2\phi)}}
{\omega_n^2} \theta_c^3 \right], 
\end{eqnarray}
where $\omega_n=\pi T (2n-1)$.
It is very important to notice 
that the leading order correction is of the order of 
$\theta_c^3$.
The correction of the linear order in $\theta_c$ cancels out 
in analogy with  Anderson theorem in the case of 
s-wave superconductor with non-magnetic impurity.
There is no quadratic order in $\theta_c$ because of the symmetry 
of the order parameter, $\cos{(2\phi)}$.
Thus, the transition temperature in the presence of doping-induced
impurity is determined by
\begin{equation}
T_c = T_{c0} - \frac{\pi n_i u_0^2 }{4} 
\left( \frac{A(k_F)}{B(k_F)} \right) \theta_c^3, 
\label{Tc}
\end{equation}
where
\begin{eqnarray}
A(k_F) &=&  \int^{\pi/2}_{0} d\phi  \left[ 
N(\phi)^2 \cos^2{(2\phi)}  \right.
\nonumber\\
& &   \left. + N(\phi) \frac{d N(\phi)}{d\phi} \cos{(2\phi)} \sin{(2\phi)}
\right] ,
\nonumber\\
B(k_F) &=&  \int^{\pi/2}_{0} d\phi N(\phi) \cos^2{(2\phi)}.
\label{ab}
\end{eqnarray}
%
We have to emphasize 
the fact that the leading order correction in cubic
order of $\theta_c$ does {\it not} depend on the shape of the
Fermi surface. This is a very robust feature due to 
symmetry of the order parameter.
However, the coefficients $A$ and $B$ depend on the shape
of the actual Fermi surface through the local density of states.
To evaluate the change of $T_c$, we have to compute
$A$ and $B$ in Eq. (\ref{ab}).
The expressions for $A$ and $B$ were obtained by expanding
the integrand of the self energy in Eq. (\ref{self}), to
quadratic order in $\theta_c$.
This is valid because scattering is finite
only for a small range of the angle $\phi$. 

Let us first estimate the reduction of $T_c$ using our
model circular Fermi surface.
The scattering rate, $\theta_c (= 1/(d k_F))$, and local density of states
are independent of angle.
In this case, the change of $T_c$ is  equal to $(\pi/2) (1/\tau) \theta_c^2$.
Taking $1/\tau=200 K$ which  would lower $T_c$ down to zero temperature
in the momentum-independent Born limit, the doping-induced
disorder lowers $T_c$ by about $6 \%$ only for $a/d=1/3$.

To extract more practical value of the reduction of $T_c$, 
let us use the scattering rate observed by ARPES data
and assume that it is proportional
to the local density of state, $N(\phi)$. 
Also $\theta_c$ is independent of angle.
The scattering rate observed by ARPES behaves as
\begin{equation} 
\frac{1}{\tau(\phi)} = \Gamma_1 \cos^2{(2\phi)} + \Gamma_0,
\end{equation}
where $\Gamma_0$ can be obtained from ARPES data.
The minimum scattering rate at the node is about $150 K$
and the maximum is about $450 K$, 
if we extract zero-temperature offset from Ref. \cite{valla2}.
Therefore, we choose $\Gamma_0=150 K$ and $\Gamma_1=300 K$.
Taking  $\theta_c \sim 1/(d k_F) \sim (5 a)/(4 \pi d)$,
we obtain
\begin{equation}
\frac{T_{c0}-T_c}{T_c} = 0.03 \sim  0.1   
\label{actualtc}
\end{equation}
where we use $T_c=91K$ and  $\frac{d}{a}=2 \sim 4$.
%
%
%
%
%
%
%
As we shown here, the effect of  small angle scattering due to 
doping-induced impurity on $T_c$ is rather small for both cases.
%

It was discussed in \cite{varma2} that the transport scattering
rate would be in cubic order of $\theta_c$. 
Our result implies that the impurities 
have small effect on both $T_c$ and the transport scattering rate
in an exactly similar manner, 
while it has rather large effect on the single
particle scattering rate.
Therefore, invoking the basic features of the cuprates, we
explain that the large scattering rate shown in ARPES 
does not contradict with the clean behavior in the transport
and the small effect on the transition temperature. 

Let us discuss the difference between the impurity
in the CuO$_2$ plane (such as Zn or Ni) and the impurity induced by doping.
The impurity induced by doping resides between the 
layers while Zn or Ni impurities go into the CuO$_2$ planes by replacing 
Cu sites. 
It has been well established that small concentration of transition metal ions
which replace Cu sites can reduce the transition
temperature substantially and shows the insulating behavior
in in-plane resistivity.\cite{gang,note0}
It is clear that the doping-induced disorder should
have stronger effect on out-of plane resistivity,
because the scattering is not restricted by small range of 
angle, and could lead insulating behavior
in out-of-plane resistivity.\cite{amplitude}
%
%
However, 
the nature of pseudogap behavior in the CuO$_2$ plane
observed in many different experiments\cite{ono} 
and the hopping matrix elements between the CuO$_2$ layers 
through the apical O\cite{anderson,atkinson}
should be also considered
for a complete analysis of the transport,
which is beyond the scope of the present paper.


To summarize, we investigated the effect of impurity induced by
doping on the transition temperature in high $T_c$ cuprates.
Since the impurity induced by doping resides between CuO$_2$ planes,
small momentum transfer is dominant in the
scattering process. 
This can be well described by the elastic scattering potential
which is finite only for a small range of the 
angle, $\theta_c \sim 1/(d k_F)$.
By computing the single particle Green function renormalized
by the impurity scattering and solving the gap equation self consistently,
we found that the leading correction on the transition
temperature is in cubic order of $\theta_c$.
The effect of doping-induced disorder on $T_c$
shows the same effect as the transport
scattering rate discussed in \cite{varma2}.
Therefore, the reduction of $T_c$ is related to the transport
scattering rate, not the single particle scattering rate.
Using the actual value of the scattering rate observed by ARPES
\cite{valla2}, we estimated the change of the transition temperature.
The reduction of the transition temperature for the optimally Bi2212
is about  $3 \sim 10\%$, while momentum-independent weak scattering
(nonmagnetic impurity in CuO$_2$ plane) could lower $T_c$ down to 
zero temperature.
Therefore, we conclude that the large elastic scattering
shown in ARPES does not contradict with either
no evidence of strong reduction of $T_c$ 
or the small effect of impurity on the transport.
%



\acknowledgements
I thank G. Sawatzky, Young S. Lee, Z. Wang, and
M. Norman for helpful discussion,  S. H. Simon for pointing
out the angle dependence of the Fermi velocity, and
C. M. Varma for motivating this investigation.
I also thank Aspen Center for Physics where this work was mostly carried out,
and ITP University of California, Santa Barbara for hospitality.
The work was conducted under the auspices of the Department
of Energy, supported (in part) by funds provided by the University of
California for the conduct of discretionary research by Los Alamos
National Laboratory.

\end{multicols}

%



\begin{references}
\bibitem{note}  Here we consider only  the hole-doped cuprates. 
So 'doping' should refer to 'hole-doping' in the entire paper.
\bibitem{note0} Here we consider Zn impurity as an elastic scattering potential,
and do not address other effect such as Kondo effect due to Zn impurity.
\bibitem{amplitude} The amplitude of scattering potential, $u_0$
can be dramatically different for the different impurity. 
The empirical fact shows that Zn has the largest
effect on $T_c$. Therefore, one should bear in mind that the effect of O (or Sr)
on $c$ axis resistivity should be quantitatively
different from that of Zn (or Ni) on
$ab$ plane resistivity.
\end{references}
\end{document}